\documentclass[a4paper]{revtex4}
\usepackage{graphicx}
\usepackage{fancyhdr}
\usepackage{amsmath}
\usepackage{color}
\usepackage{amsmath, amssymb, slashed, epsf}
\usepackage{graphicx}
\usepackage{cancel}
\usepackage{array}

\begin{document}

\title{Probing composite Higgs models by measuring phase shifts at LHC}

%

\author{K.~Kaneta}
\affiliation{ICRR, the University of Tokyo, Kashiwa, Chiba 277-8582, Japan}

\begin{abstract}
Composite Higgs models are an attractive scenario, where the discovered Higgs boson is regarded as a Nambu-Goldstone boson associated with spontaneous breakdown of a global symmetry of more fundamental theory. This class of models predicts violation of perturbative unitarity at high energies, and new resonances are expected to appear around TeV scale to maintain the unitarity, while a sizable phase shift is predicted in certain scattering amplitude. We investigate the new resonance scale from the phase shift by drawing analogies with pion physics in QCD. The detectability of the phase shift at LHC and the ILC is also discussed.
This talk was given in {\it HPNP 2015} at University of Toyama and based on the work in collaboration with S.~Kanemura, T.~Shindou and N.~Machida (arXiv:1410.8413 [hep-ph]).
\end{abstract}

\maketitle

\thispagestyle{fancy}

\section{Introduction}
\label{sec:Introduction}

The observation of the Higgs boson with the mass around 125 GeV~\cite{Aad:2012tfa} is a great achievement at LHC although we could not find any signals of new physics beyond the standard model (SM).
On the other hand, the observed value of the Higgs boson mass is a mystery in itself, and we have been plagued by the problem known as {\it naturalness} of the electroweak scale.
Since the Higgs boson mass is a dimension-full parameter, we come up with a question: why is the Higgs boson mass around 125 GeV against large quantum corrections?
To answer this question, we may need a new paradigm, where a promising candidate is supersymmetry.
As an alternative, strong dynamics is an attractive scenario in which the Higgs boson is regarded as a composite state.
In this article we will focus on this possibility.

An analogy with pion physics in QCD might be helpful to investigate the composite Higgs scenario.
In the meson spectrum in QCD (neutral) pion is the lightest scalar particle since it is a pseudo Nambu-Goldstone boson (pNGB) associated with the chiral symmetry breaking.
Regarding the particle spectrum in the SM, the Higgs boson is the lightest scaler particle.
If we draw an analogy with the pion, the Higgs boson could be identified as a pNGB associated with spontaneous breakdown of a global symmetry of more fundamental theory.
In this point of view new resonance states are expected to appear above the Higgs mass scale as is the case with the rho mesons in QCD.
By remembering the broad width of the rho mesons, phase shifts of scattering amplitudes can be observed in the tail region of the rho meson mediated cross sections even if the scattering energy does not reach the resonance peak~\cite{Hyams:1973zf}.
The same measurement can be applied in the composite Higgs scenario if the second lightest resonance above the Higgs mass scale has a relatively broad decay width.

\section{Phase shift measurement at LHC}
\label{sec:LHC}

Although direct observations of new particles can be a strong evidence of new physics, phase shift measurements would be useful in the case that new resonances have a broad width and collision energies are not sufficient to directly produce them.
In this section we discuss a possibility to observe the phase shift at LHC.
The expected new resonance couples to the Higgs doublet field, and thus, the phase shift would appear in production processes of the longitudinally polarized weak gauge bosons.
However, it is challenging to observe the longitudinal mode at LHC~\cite{Contino:2010mh}.
We alternatively utilize an interference effect between the longitudinal mode and the production modes of transversely polarized gauge bosons~\cite{Keung:2008ve,Cao:2009ah,Murayama:2014yja}.

Let us focus on the process $pp\to W^+Z\to l^+\nu l^+ l^-$ at LHC.
In this process the squared amplitude can be written by
\begin{eqnarray}
	\left|\sum_{\lambda_W,\lambda_Z}
	{\cal M}_{\rm Prod}^{u\bar d\to W^+Z}(\theta;\lambda_W,\lambda_Z)
	{\cal M}_{\rm Decay}^{W^+\to l^+\nu}(\theta_1,\phi_1;\lambda_W)
	{\cal M}_{\rm Decay}^{Z\to l^+l^-}(\theta_2,\phi_2;\lambda_Z)
	\right|^2,
\end{eqnarray}
where ${\cal M}_{\rm Prod}^{u\bar d\to W^+Z}(\theta;\lambda_W,\lambda_Z)$ is the production amplitude as a function of $\theta$, $\lambda_W$ and $\lambda_Z$ being the scattering angle, the polarization of $W$ and that of $Z$, respectively.
${\cal M}_{\rm Decay}^{W^+\to l^+\nu}$ and ${\cal M}_{\rm Decay}^{Z\to l^+l^-}$ are decay amplitudes of $W$ and $Z$ bosons, respectively, where $\theta_{1 (2)}$ and $\phi_{1 (2)}$ are the polar and azimuthal angles of the final state leptons coming from $W (Z)$ decay, respectively.
As already mentioned, a phase shift can be induced in the production amplitude for $(\lambda_W, \lambda_Z)=(0,0)$, and thus we parametrize the phase shift $\delta$ by ${\cal M}_{\rm Prod}^{u\bar d\to W^+Z}(\theta;0,0)\to {\cal M}_{\rm Prod}^{u\bar d\to W^+Z}(\theta;0,0)e^{i\delta}$.
On the other hand, the decay amplitudes are proportional to ${\cal M}_{\rm Decay}^{W^+\to l^+\nu}(\theta_1,\phi_1;\lambda_W)\propto e^{i\lambda_W\phi_1}$ and ${\cal M}_{\rm Decay}^{Z\to l^+l^-}(\theta_2,\phi_2;\lambda_Z)\propto e^{i\lambda_Z\phi_2}$.
We therefore obtain the term proportional to $\delta$ through the interference term: $|A e^{i\delta}+B e^{i(\lambda_W\phi_1-\lambda_Z\phi_2)}|^2\supset\sin(\lambda_W\phi_1-\lambda_Z\phi_2)\sin\delta$ where $A$ and $B$ are some coefficients depending on other parameters.
It turns out that the non-vanishing $\delta$ can be probed by $\phi_1$ and/or $\phi_2$ dependence of the final state leptons.
It should be emphasized that, when we detect the final state charged lepton coming form $W$ decay, it is possible to measure $\phi_1$ dependence induced by non-vanishing $\delta$ even in the case of $(\lambda_W,\lambda_Z)\neq (0,0)$.

\begin{figure}[tbp]
	\begin{minipage}[t]{0.5\columnwidth}
		\begin{center}
			\includegraphics[width=0.8\columnwidth]{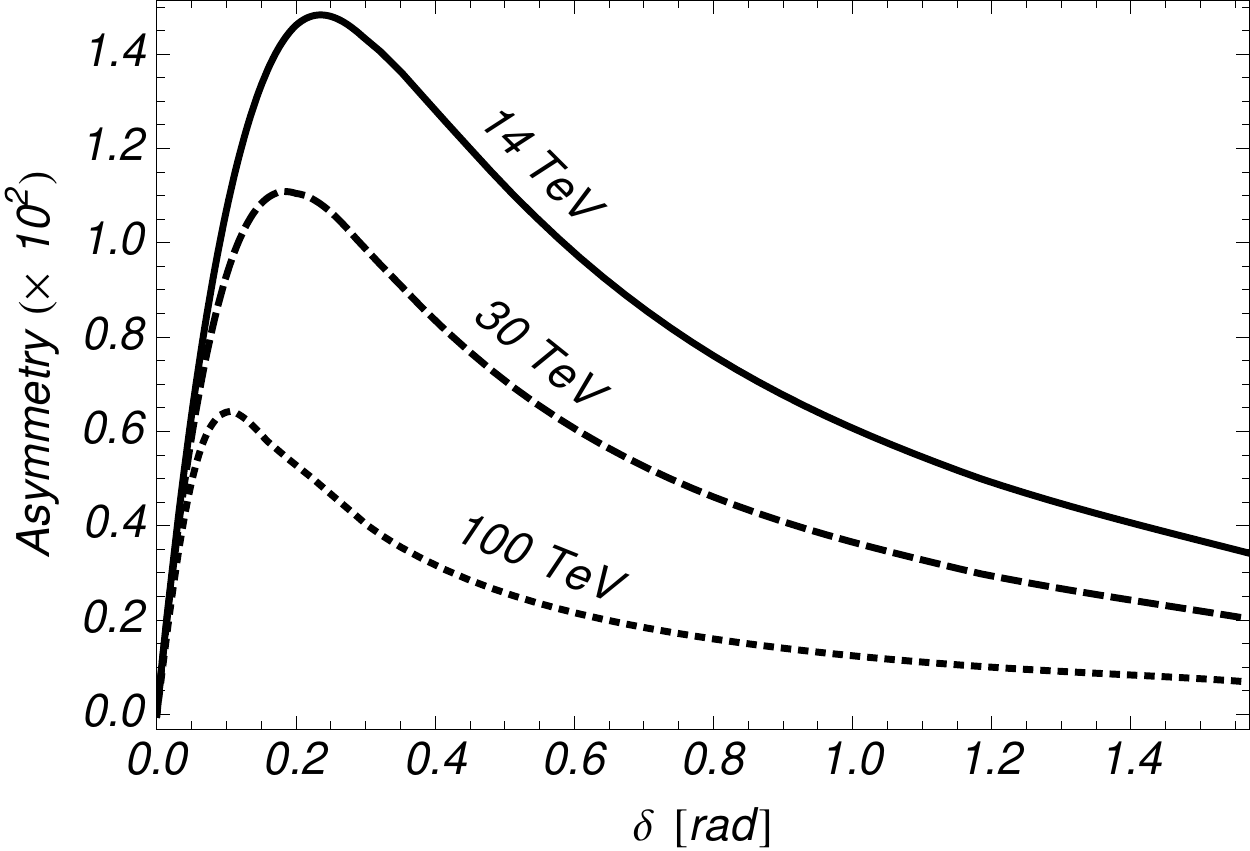}
		\end{center}
	\end{minipage}%
	\begin{minipage}[t]{0.5\columnwidth}
		\begin{center}
			\includegraphics[width=0.8\columnwidth]{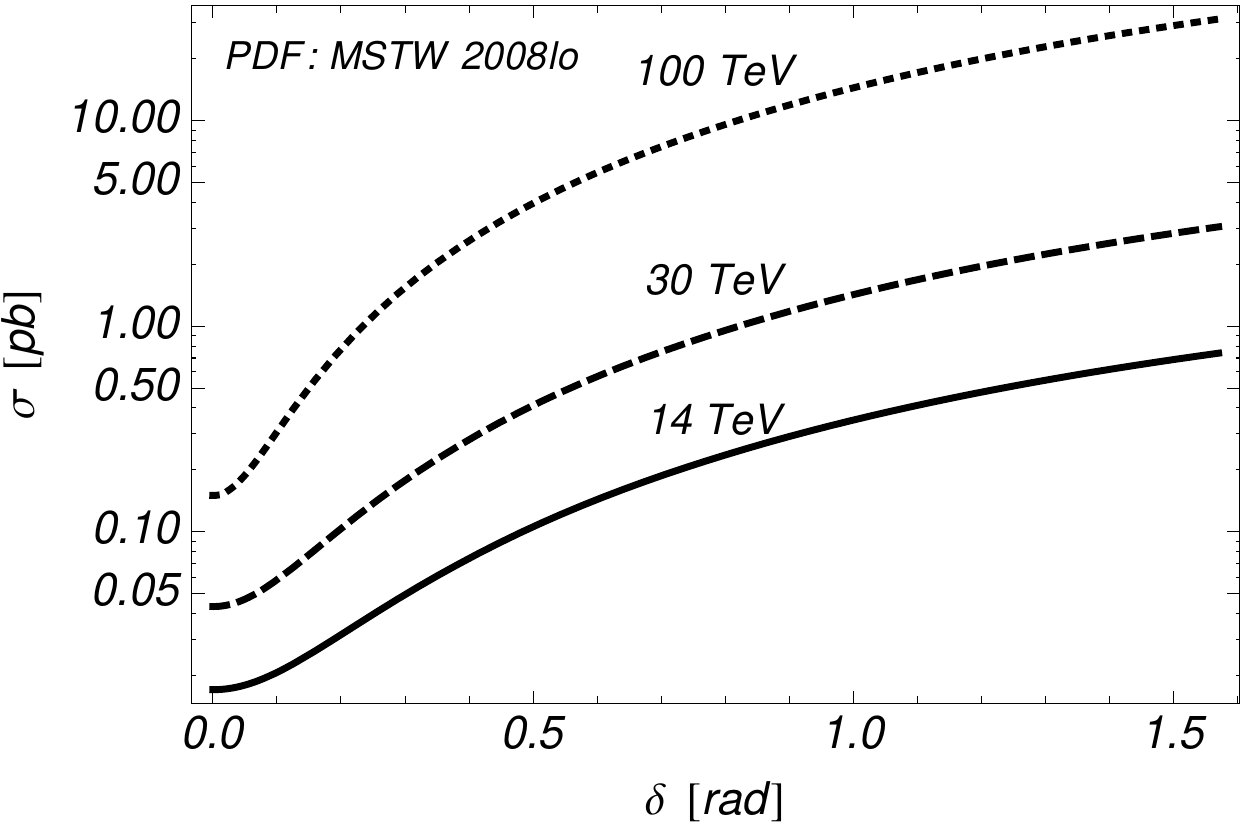}
		\end{center}
	\end{minipage}
	\caption{\sl\small {\bf Left panel:} Asymmetry $A_\pm$ as a function of phase shift $\delta$. The solid, dashed and dotted lines show the cases that the collision energy is set to be 14 TeV, 30 TeV and 100 TeV, respectively. {\bf Right panel:} Total cross section $\sigma$ as a function of phase shift $\delta$. The meaning of the lines is the same as the left panel.}
	\label{fig:pp}
\end{figure}

To see $\phi_1$ dependence, we utilize the quantity defined by
\begin{eqnarray}
	A_\pm = |\sigma_+ - \sigma_+|/(\sigma_+ + \sigma_-),~~~~~
	\sigma_\pm\equiv \sigma(\sin\phi_1\gtrless 0),
\end{eqnarray}
where $\sigma_{+(-)}$ is the cross section for the case that the charged lepton coming from $W$ decay goes to "above (below)" the production plane.
To avoid the misidentification of $u$-direction, the cross sections are given by integrating over $0<\cos\theta<1$~\footnote{For more detailed discussion, see Refs.~\cite{Murayama:2014yja,Kanemura:2014kga}.}.
The left panel of Fig.~\ref{fig:pp} shows the asymmetry $A_\pm$ as a function of phase shift $\delta$ at $pp$ colliders.
The solid, dashed and dotted lines show the cases that the collision energy is set to be 14 TeV, 30 TeV and 100 TeV, respectively.
In the case of 14 TeV LHC the maximal asymmetry $A_\pm\sim 1.5~\%$ appears around $\delta\sim 0.24$.
The total cross section $\sigma$ is then given by $\sigma\sim 3.7~{\rm fb}$ as shown in the right panel of the figure.
When we have 300 fb$^{-1}$ of the integrated luminosity, the statistical error is about 3 \%, and it is difficult to measure the asymmetry in this case.
If we have, on the other hand, 3000 fb$^{-1}$, the statistical error could be less than the asymmetry, and thus it might be possible to observe the asymmetry.

\section{Phase shifts and new resonance scales}
\label{sec:unitarity}

Once we obtain the information of the phase shift by collider experiments, we can anticipate a new resonance scale under some assumptions.
Here let us discuss a relation between the phase shift $\delta$ and new resonance scale parametrized by its mass $m_\rho$ and decay width $\Gamma_\rho$.
In this section we concentrate on so-called minimal composite Higgs model (MCHM)~\cite{Agashe:2004rs}.
This model is based on global $SO(5)$ symmetry which is spontaneously broken into $SO(4)$, and four NGBs are identified as the SM Higgs doublet.
In this model the gauge interaction of the Higgs field is deviated form that of the SM prediction, which is given by $g_{hVV}=g_{hVV}^{\rm SM}\sqrt{1-\xi}$ with $g_{hVV}^{\rm SM}$ being the $h$-$V$-$V$ ($V=W,Z$) coupling in the SM.
The parameter $\xi$ is defined by $\xi\equiv v_{\rm ew}^2/f^2~(<1)$, where $v_{\rm ew}$ and $f$ are vacuum expectation value of the Higgs field and the breaking scale of $SO(5)$, respectively.
Although there are many variations of MCHM according to matter representations~\footnote{A comprehensive list of models and Higgs boson couplings is shown in Ref.~\cite{Kanemura:2014kga}}, the gauge interaction of the Higgs field is completely determined by the coset space of $SO(5)/SO(4)$, and thus it does not depend on the matter sector.

The deviation of the $h$-$V$-$V$ coupling is responsible for the violation of perturbative unitarity at high energies, and lets a partial wave amplitude $a_l$ $(l=0,1,2,\cdots)$ go outside the unitarity circle  defined by
\begin{eqnarray}
 	{\rm Re}[a_l]^2 + ({\rm Im}[a_l]-1/2)^2\leq (1/2)^2.
 	\label{eq:unitarity circle}
 \end{eqnarray} 
 For example, $s$-wave amplitude of elastic $WW$ scattering is given by
 \begin{eqnarray}
 	a_0 = G_F\xi S/(16\sqrt{2}\pi) + G_F(m_h^2 - M_W^2)(1-\xi)/(4\sqrt{2}\pi),
 	\label{eq:s-wave amp}
 \end{eqnarray}
where $S$, $G_F$, $m_h$ and $M_W$ are squared center-of-mass energy, Fermi constant, the Higgs boson mass and $W$ boson mass, respectively.
The amplitude of Eq.~(\ref{eq:s-wave amp}) linearly depends on $S$, and thus exceeds the unitarity limit for non-vanishing $\xi$ at some high energies.

In the composite Higgs picture, on the other hand, new resonance states are expected to appear below the scale of unitarity violation, and (at least partly) maintain the unitarity of the amplitudes by providing non-vanishing phase shifts (or equivalently imaginary part of the amplitudes).
In the elastic $s$-wave scattering case the imaginary part of $a_0$ can be written in terms of the real part of $a_0$ since the amplitude is on the unitarity circle.
We here parametrize the phase shift by $\delta\equiv\tan^{-1}({\rm Im}[a_0]/{\rm Re}[a_0])$, and thus obtain
\begin{eqnarray}
	\delta = \tan^{-1}\left[1/(2{\rm Re}[a_0] \pm \sqrt{1/(2{\rm Re}[a_0])^2-1}\right]
	\label{eq:delta from a0}
\end{eqnarray}
by substituting the equality of Eq.~(\ref{eq:unitarity circle}), where ${\rm Re}[a_0]$ is given by Eq.~(\ref{eq:s-wave amp}).

\begin{figure}[tbp]
	\begin{center}
		\includegraphics[width=0.35\columnwidth]{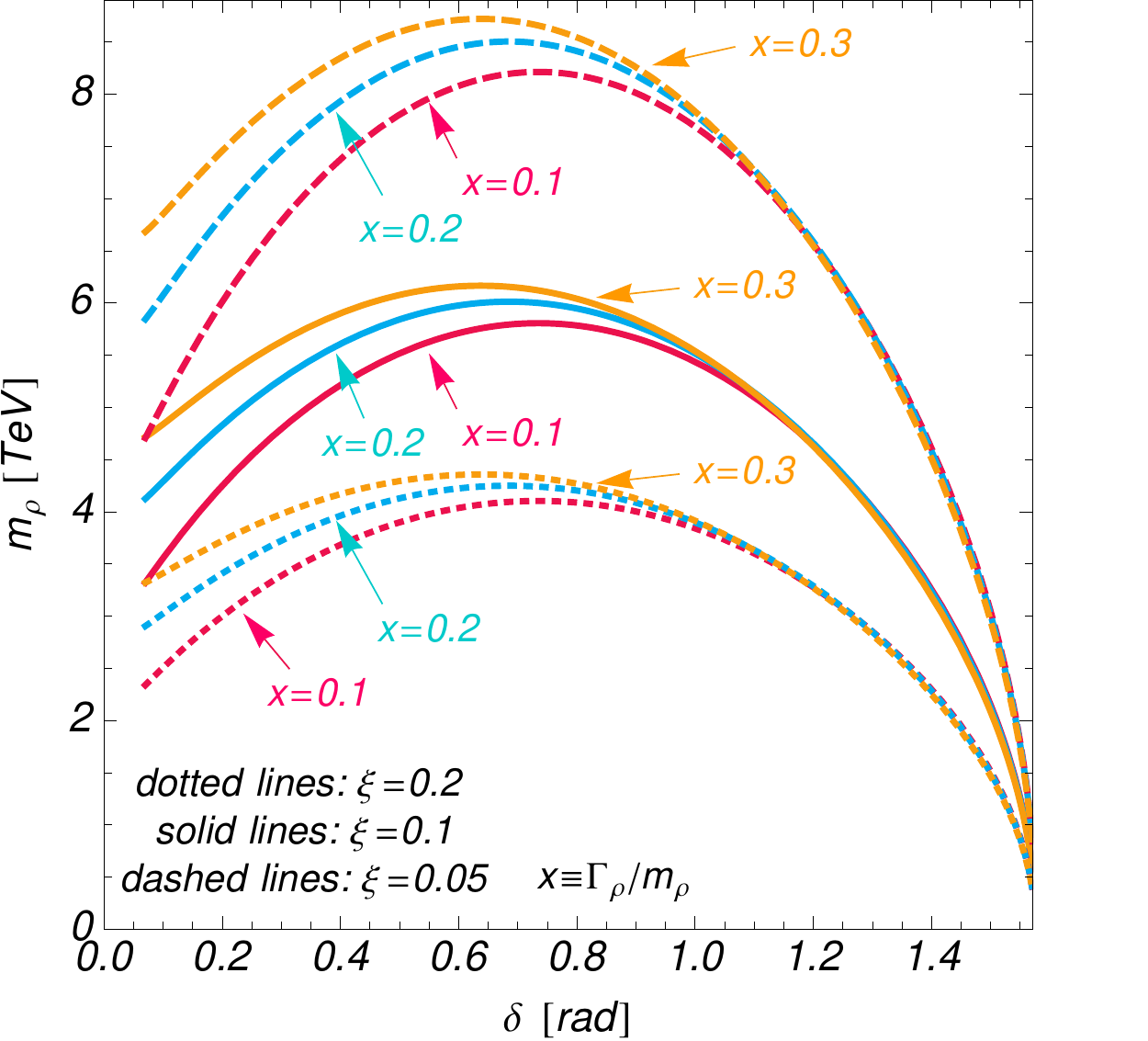}
	\end{center}
	\caption{\sl\small New resonance scale $m_\rho$ as a function of $\delta$ by fixing $\xi$ and $x\equiv \Gamma_\rho/m_\rho$, where we assume that Eqs.~(\ref{eq:delta from a0}) and (\ref{eq:delta from VMD}) are the same.}
	\label{fig:mass}
\end{figure}

We again draw an analogy with pion physics in which the phase shift has been experimentally observed~\cite{Hyams:1973zf,Protopopescu:1973sh}.
It is known that the observed $\delta$ can be well fitted by
\begin{eqnarray}
	\delta =
	\left\{
	\begin{array}{l}
		\tan^{-1}\left[(\Gamma_\rho)S/(m_\rho^2+\Gamma_\rho^2-S)\right]~~~~~~~~\text{for}~~~~~S<m_\rho^2+\Gamma_\rho^2\\
		\tan^{-1}\left[(\Gamma_\rho)S/(m_\rho^2+\Gamma_\rho^2-S)\right]+\pi~~~\text{for}~~~~~S\geq m_\rho^2+\Gamma_\rho^2
	\end{array}
	\right.
	\label{eq:delta from VMD}
\end{eqnarray}
so that the phase shift is parametrized by $m_\rho$, $\Gamma_\rho$ and $S$~\cite{Murayama:2014yja}.
We here put an assumption that the required $\delta$ from perturbative unitarity shown in Eq.~(\ref{eq:delta from a0}) has the same shape as the observed $\delta$ in pion scatterings, which is given by Eq.~(\ref{eq:delta from VMD}).
Under this assumption, $m_\rho$ and $\Gamma_\rho$ now represent the mass scale and the decay width of the expected new resonance which maintains perturbative unitarity.
We therefore obtain a relation among $\xi$, $m_\rho$ and $\Gamma_\rho$ by eliminating $S$ in Eqs.~(\ref{eq:delta from a0}) and (\ref{eq:delta from VMD}), which is shown in Fig.~\ref{fig:mass} on $\delta$-$m_\rho$ plane by fixing $\xi$ and the ratio of $\Gamma_\rho/m_\rho$.
Regarding the behavior of the lines in Fig.~\ref{fig:mass}, lower $m_\rho$ is predicted for larger $\xi$ since the scale of unitarity violation decreases for large $\xi$.
Once $\xi$ is fixed, lower $m_\rho$ is predicted for smaller $\Gamma_\rho$ since the resonance cannot provide sufficient phase shift before unitarity is violated.

\section{Phase shift measurement at the ILC}
\label{sec:ILC}

\begin{figure}[tbp]
	\begin{minipage}[t]{0.5\columnwidth}
		\begin{center}
			\includegraphics[width=0.9\columnwidth]{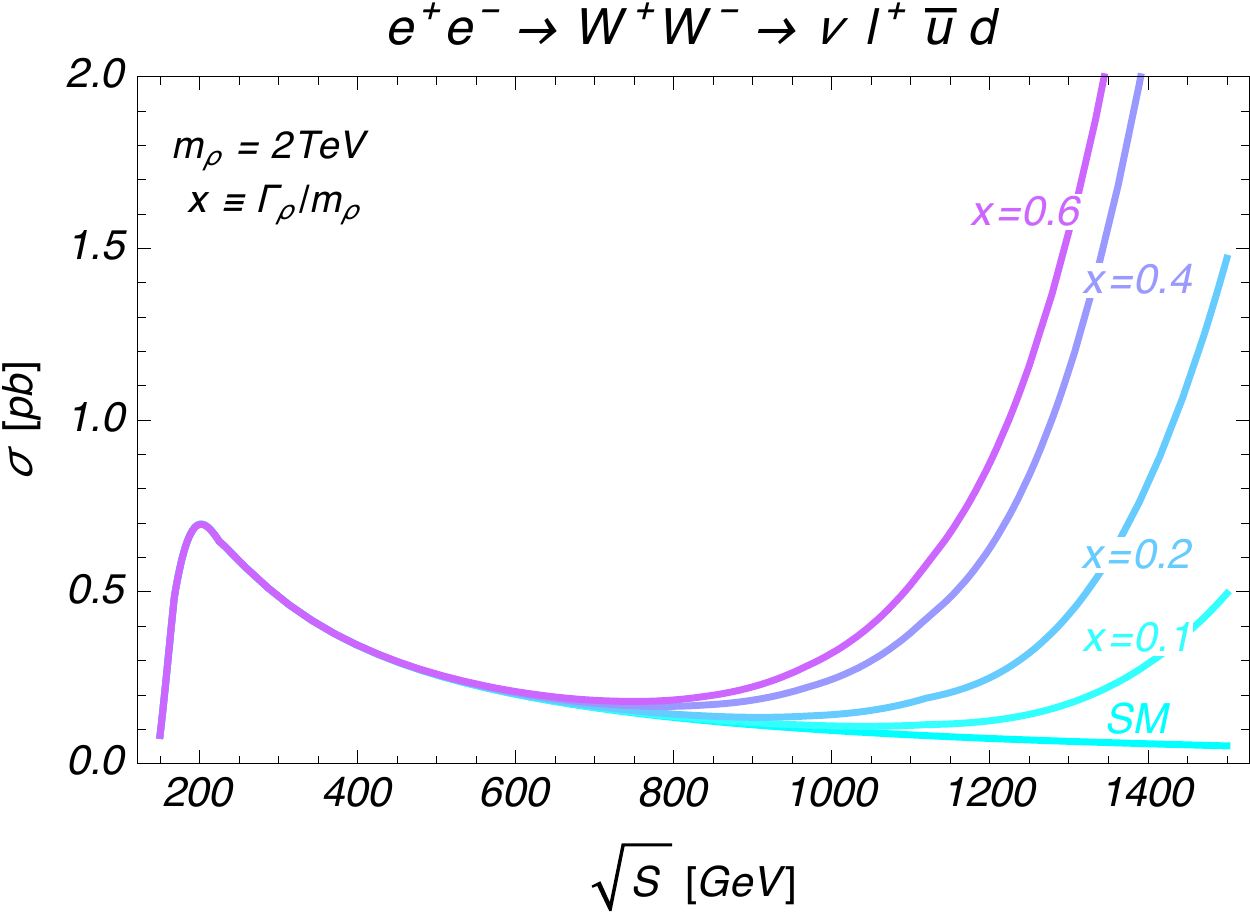}
		\end{center}
	\end{minipage}%
	\begin{minipage}[t]{0.5\columnwidth}
		\begin{center}
			\includegraphics[width=0.55\columnwidth]{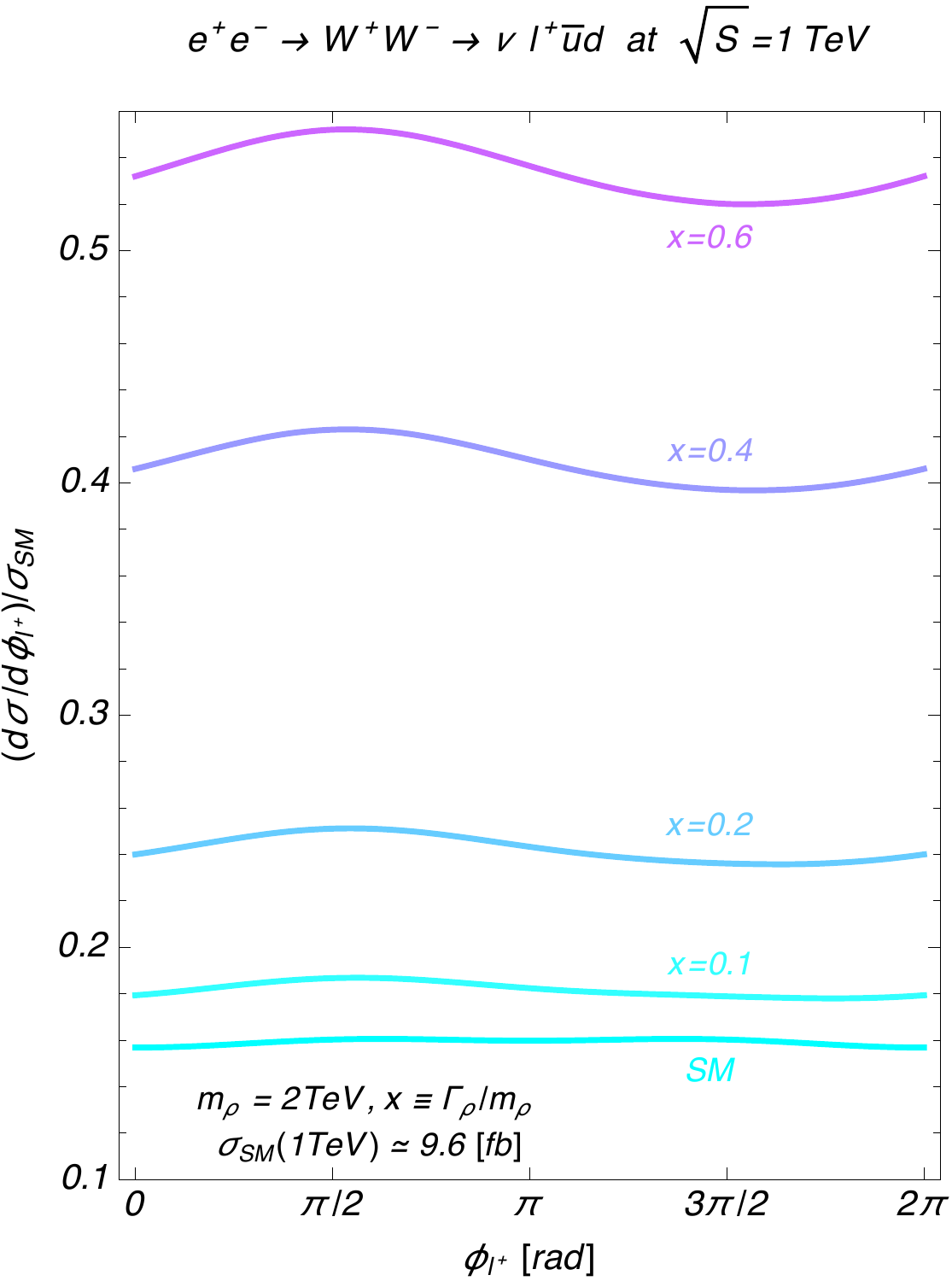}
		\end{center}
	\end{minipage}
	\caption{\sl\small {\bf Left panel:} Total cross section $\sigma$ as a function of center-of-mass energy $\sqrt{S}$, where we assume that the phase shift $\delta$ follows Eq.~(\ref{eq:delta from VMD}), and take the mass of resonance 2 TeV. {\bf Right panel:} Azimuthal angle distribution normalized by the SM cross section $\sigma_{SM}\simeq 9.6$ fb at 1 TeV collision.}
	\label{fig:ee}
\end{figure}

Finally let us discuss phase shift measurement at the International Linear Collider (ILC)~\cite{KKSM}.
Here we focus on the process $e^+e^-\to W^+W^-\to l\nu q \bar q'$ where the produced $W$ bosons decay in semi-leptonic way.
The method to measure phase shifts is the same as in the case of LHC, in which azimuthal angle dependence of the final state leptons reflects a non-vanishing phase shift~\cite{Barklow:1997nf}.
The left panel of Fig.~\ref{fig:ee} shows total cross section $\sigma$ as a function of center-of-mass energy $\sqrt{S}$, where we assume that the phase shift $\delta$ follows Eq.~(\ref{eq:delta from VMD}), and take the mass of resonance 2 TeV.
For example, in the case of $\Gamma_\rho/m_\rho=0.1$ the total cross section at 1 TeV collision is about 10 \% larger than the SM prediction.
The right panel of Fig.~\ref{fig:ee} shows the azimuthal angle distribution of the final state charged lepton at 1 TeV collision.
In the case of $\Gamma_\rho/m_\rho=0.1$, for example, the difference between the maximal and minimal values is $O(1)$ \%, and thus there is a possibility to observe the phase shift if the cross section and the angular distribution can be measured with $O(1)$ \% level.

\section{Summary}
\label{sec:summary}

We have discussed that phase shift measurements can be an important method to search new resonance states, which are expected in the composite Higgs scenario, at collider experiments.
There, we have drawn analogies with pion physics in QCD, and tried to give a relation among $\delta$, $\xi$, $m_\rho$ and $\Gamma_\rho$ by utilizing perturbative unitarity.
More detailed discussions can be found in Ref.~\cite{Kanemura:2014kga,KKSM}.

\begin{acknowledgments}

This talk is based on Ref.~\cite{Kanemura:2014kga} and partly consists of a work in progress in collaboration with S.~Kanemura, T.~Shindou and N.~Machida~\cite{KKSM}.
K.K. expresses thanks to the organizers of the conference {\it HPNP 2015} at University of Toyama for their warm hospitality.

\end{acknowledgments}

\bigskip 

\end{document}